\documentclass[9pt,journal,compsoc]{IEEEtran}
\usepackage{comment}
\usepackage{hyperref}
\usepackage{url}
\usepackage{flushend}
\usepackage{enumitem}	
\usepackage{amssymb}
\usepackage{xcolor}
\usepackage{pgf,interval}
\usepackage{ifthen}
\usepackage{booktabs}
\usepackage{tikz}
\usepackage{mathtools}
\usepackage[flushleft]{threeparttable}
\usepackage{multirow}
\usepackage{multicol}
\usepackage[flushleft]{threeparttable} 

\usepackage{amssymb}

\usepackage{hyphenat}

\usepackage{graphicx}
\usepackage{csquotes}
\usepackage{float}

\usepackage{balance}
\usepackage[binary-units=true]{siunitx} 
\usepackage{subfig}
\ifCLASSOPTIONcompsoc
  \usepackage[nocompress]{cite}
\else
  \usepackage{cite}
\fi

\begin{document}
\bstctlcite{IEEEexample:BSTcontrol}

\title{DNA Pre-alignment Filter using Processing Near Racetrack Memory}

\author{Fazal~Hameed,
        Asif~Ali~Khan,
        Sebastien~Ollivier,
        Alex~K.~Jones~\IEEEmembership{Senior~Member,~IEEE},
        and~Jeronimo~Castrillon,~\IEEEmembership{Senior~Member,~IEEE}
\IEEEcompsocitemizethanks{\IEEEcompsocthanksitem F. Hameed is with the IST, Islamabad 44000, Pakistan.
\IEEEcompsocthanksitem S. Ollivier and A. K. Jones are with the University of Pittsburgh, 1238 Benedum Hall, 3700 O’Hara Street, Pittsburgh, PA 15261, USA.
\IEEEcompsocthanksitem A. A. Khan and J. Castrillon are  with the Chair for Compiler Construction, Technische Universit\"at Dresden, 01069 Dresden, Germany.\protect\\
E-mails: \{fazal.hameed,asif\_ali.khan, jeronimo.castrillon\}@tu-dresden.de, \{sbo15,akjones\}@pitt.edu.
}
}


\markboth{}%
{Hameed \MakeLowercase{\textit{et al.}}: Prealignment filtering using Processing near Racetrack Memory}

\IEEEtitleabstractindextext{%
\begin{abstract}
Recent DNA \emph{pre-alignment} filter designs employ DRAM for storing the reference genome and its associated meta-data.
However, DRAM incurs increasingly high energy consumption background and refresh energy as devices scale.
To overcome this problem, this paper explores a design with \emph{racetrack memory} (RTM)--an emerging non-volatile memory that promises higher storage density, faster access latency, and lower energy consumption. 
Multi-bit storage cells in RTM are inherently sequential and thus require data placement strategies to mitigate the performance and energy impacts of shifting during data accesses. 
We propose a near-memory pre-alignment filter with a novel data mapping and several shift reduction strategies designed explicitly for RTM. 
On a set of four input genomes from the 1000 Genome Project, our approach improves performance and energy efficiency by 68\% and 52\%, respectively, compared to the state of the art proposed DRAM-based architecture.

\end{abstract}

}

\maketitle

\IEEEdisplaynontitleabstractindextext

\IEEEpeerreviewmaketitle

\IEEEraisesectionheading{
\section{Introduction}\label{sec:intro}}
Sequence alignment is a fundamental but computationally expensive step in genomic analysis, where short DNA fragments of a query genome are mapped and compared against a reference genome. High-throughput sequencing (HTS) technologies generate a massive amount of sequencing data, making the sequence alignment a performance bottleneck. In particular, solutions based on dynamic programming are extremely slow and quickly become impractical~\cite{Waterman_81}. 
Current algorithms thus use a seed-and-extension approach that better scales with the data volume~\cite{bwa2, minimap2}. 
However, these algorithms still spend considerable time analyzing genome locations that eventually do not align. 
To alleviate this problem, recent research employs pre-alignment filters~\cite{GRIM_2018, sneakysnake, darwin}.
These filters significantly reduce the number of DNA fragments that are passed on to the more computational expensive alignment phase.

Recent pre-alignment filters have been implemented on GPUs~\cite{gk_gpu}, FPGAs~\cite{shouji}, ASICs~\cite{darwin} or on CPUs with die-stacked DRAMs~\cite{GRIM_2018, ALPHA_2021}.
This work mainly concentrates on the near-data processing in die-stacked architectures because it significantly reduces data movement between the processor and the memory, increasing performance compared to dedicated accelerators. 
However, the background and refresh energy consumption of DRAM main memory becomes a significant challenge as background power increases significantly as the capacity of DRAM chips increases~\cite{Bhatti, Bhatti_15}. 

A potential solution to the DRAM energy wall is to replace DRAM with emerging non-volatile memory (NVM) technologies, \textit{e.g.,} STT-RAM~\cite{STTRAM_ours}, phase change memory~\cite{PCM}, resistive RAM~\cite{ReRAM} or Racetrack memory~\cite{pieee}. Despite being energy efficient, not all NVM technologies can compete with DRAM in terms of performance.  \emph{Racetrack memory} (RTM) delivers comparable performance with significantly reduce background power and much higher density compared to DRAM~\cite{pieee}. 
RTM cells are like STT-RAM cells where the free layer has been extended to a magnetic nanowire (referred to as a track) that stores multiple bits, or \textit{domains}, separated by domain walls. 
Retrieving a particular bit from the track requires \emph{shifting} the  domain to be aligned with the access port.  Thus, RTM inherits the fast latency, high endurance, and non-volatility from STT-RAM, but its 
performance and energy efficiency depends on its shifting overhead, making data placement important.  

This paper explores \emph{filtering in racetrack memory} (FIRM), a collaboratively designed, RTM, in-die-stacked architecture for the pre-alignment filtering, retaining the CMOS logic layer from recent designs~\cite{GRIM_2018,ALPHA_2021}. Our evaluations show that naively replacing DRAM with RTM leads to significant performance and energy overheads. We propose RTM optimized storage and data management approaches that leverage characteristics of the pre-alignment filtering algorithm to significantly reduce the RTM shifts and maximize access parallelism across subarrays.  In particular we make the following contributions:
\begin{itemize}
    \item We \textit{interleave} accesses of different tokens across subarrays which reduces latency by \textit{pipelining} memory accesses.
    \item We \textit{minimize shifts} by ensuring that subarrays without a matching token will not shift.
    \item We demonstrate a \textit{preshifting} technique to avoid latency overhead of shifts.
    \item We utilize a circular \textit{unlimited single-shift} technique to allow buffers to automatically reset to their original position.
\end{itemize}
Applying interleaved accesses and minimizing shifts for tokens not in the query read decreases runtime and energy versus the state-of-the-art DRAM technique~\cite{ALPHA_2021} by 
63\% and 48\%
, respectively.  When adding preshifting and unlimited single-shiftings, these improvements grow to 68\% and 52\%, respectively.

\vspace{-.05in}
\section{Background and related work}
\vspace{-.025in}
\label{sec:related}
This section provides the background on RTM architectures and the pre-alignment filters. 
\vspace{-.1in}
\subsection{Racetrack memory}
\vspace{-.05in}
\label{subsec:RTMOrg}
A single RTM magnetic nanowire can store up to 100 data bits.  We assume planar RTM tracks that are grouped to form subarrays. 
Each track in a subarray has one or more \emph{access ports} that enable reading/writing data from/to the track. 
Due to the larger area footprint of access transistors, each track can only have a limited number of access ports. Therefore, data placement to minimize shifting for data accesses is an important performance and energy optimization criterion. 
\vspace{-.1in}
\subsection{State-of-the-art pre-alignment filters}
\vspace{-.05in}
\label{subsec:ALPHA}

In the state-of-the-art pre-alignment~\cite{GRIM_2018,ALPHA_2021}, the reference genome $R$ is divided into $n$ fixed-size bins. 
The query genome $Q$ is divided into $m$ reads where each read $q_i$ contains many nucleotides.  The four nucleotides in DNA are represented by the alphabet $\Gamma = \{A, C, G, T\}$. The matching operation is performed at the granularity of a \emph{token} which is a string over $\Gamma$ consisting of $y$ nucleotides, \textit{i.e.,} in $\Gamma^y$. 
We assume $y=5$, the minimum token size to provide high fidelity~\cite{GRIM_2018}. 
Each nucleotide can be encoded as a 2-bit symbol where $A=$``00'', $C=$``01'', $G=$``10'', and $T=$``11''.  Mathematically, ordered set of all possible size five tokens is $S = \{\Gamma^5_i \mid 0 < i < 1023 \land \Gamma^5_i <^{lex} \Gamma^5_{i+1} \}$ indexed by $f : S \mapsto i \in \mathbb{N}^{+}\cup \{0\}$. 

\begin{figure}[bp]
\vspace{-.15in}
\centering
\includegraphics[width=\linewidth]{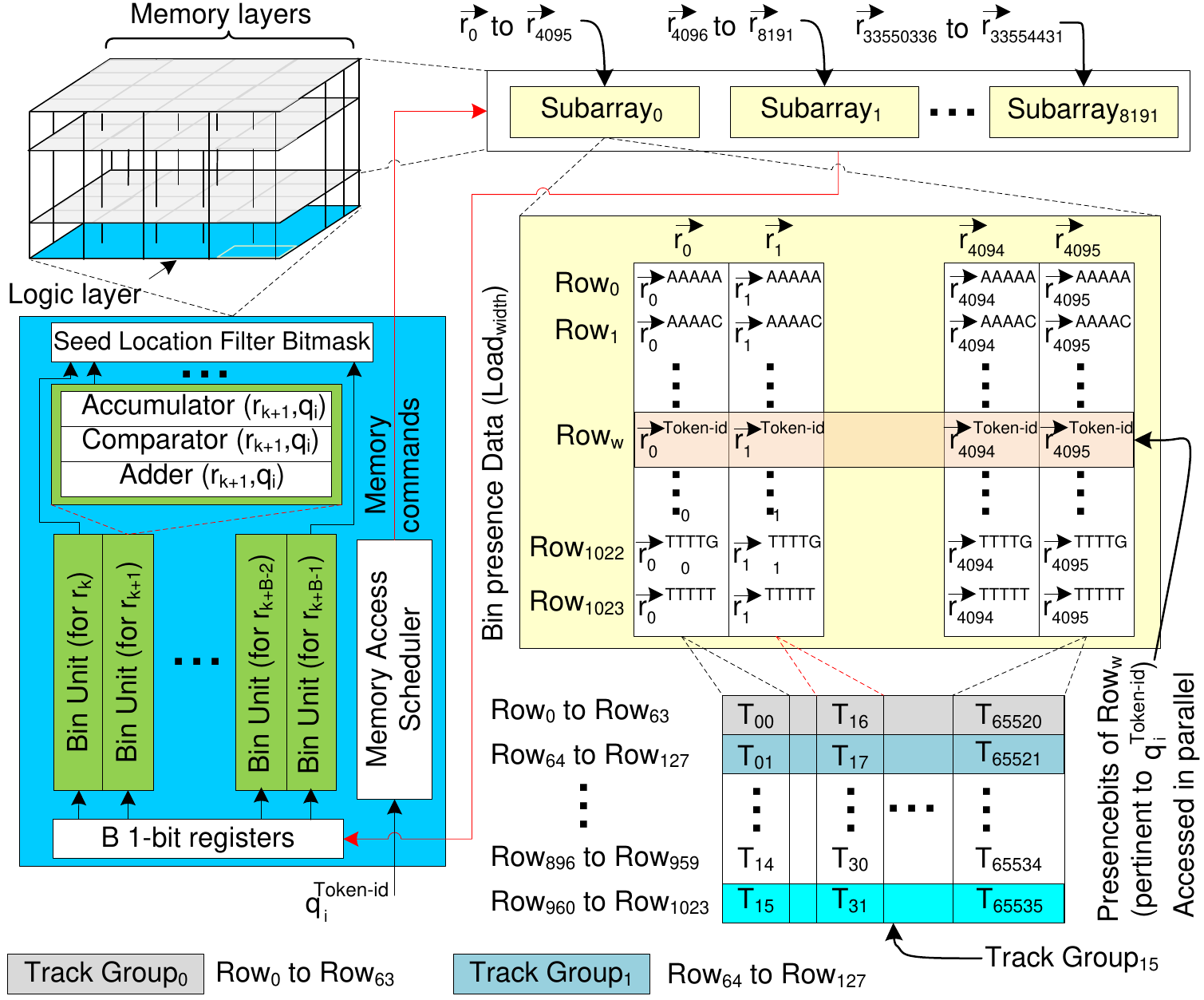}
\vspace{-.15in}
\caption{ALPHA filter hardware design and data mapping of 4096 bins (\textit{i.e.,} $r_0$ to $r_{4095}$) in $R$ to Subarray$_0$. The bottom-right of the figure shows how bit-vectors are organized in RTM tracks.} 
\label{fig:GRIM3D}
\end{figure}

For each bin $r_k \in R$ a bit-vector $\vec{r}_k$ with 1024 locations is used shown in yellow in Fig.~\ref{fig:GRIM3D}. 
Each bit in $\vec{r}_k$ is indexed by $f(\Gamma^5_i)$ which is 0 to 1023 for $AAAAA$ to $TTTTT$ as discussed above. 
If token $j$ exists in $r_k$, its corresponding bit (referred to as \emph{presence bit}) in $\vec{r}_k$ is set, \textit{i.e.,} $\vec{r}_k[j] = 1$.
In GRIM~\cite{GRIM_2018}, each read $q_i \in Q$ is matched against $r_k$ on a token-by-token basis. 
For each token in $q_i$, the relevant presence bit $\vec{r}_k[f(token)]$ is accessed and, if set, a counter $c_{k}$ is incremented.  
Once the relevant bits from $\vec{r}_k$ are accessed, the accumulator $c_{k}$ is compared to a predefined threshold $T$.
Only the bins $r_k$ are selected for the seed extension if $c_{k} > T$ and all other bins are discarded. That is, a bin $r_k$ is selected for read $q_i$ if $\sum_{token \in q_i}\vec{r}_k[f(token)] > T$. 
A near-memory accelerator architectures where the memory layers store the token presence bits of the bins in $R | \forall k, r_k \in R$. 

%

\begin{figure}[tbp]
\centering
\includegraphics[scale=0.65]{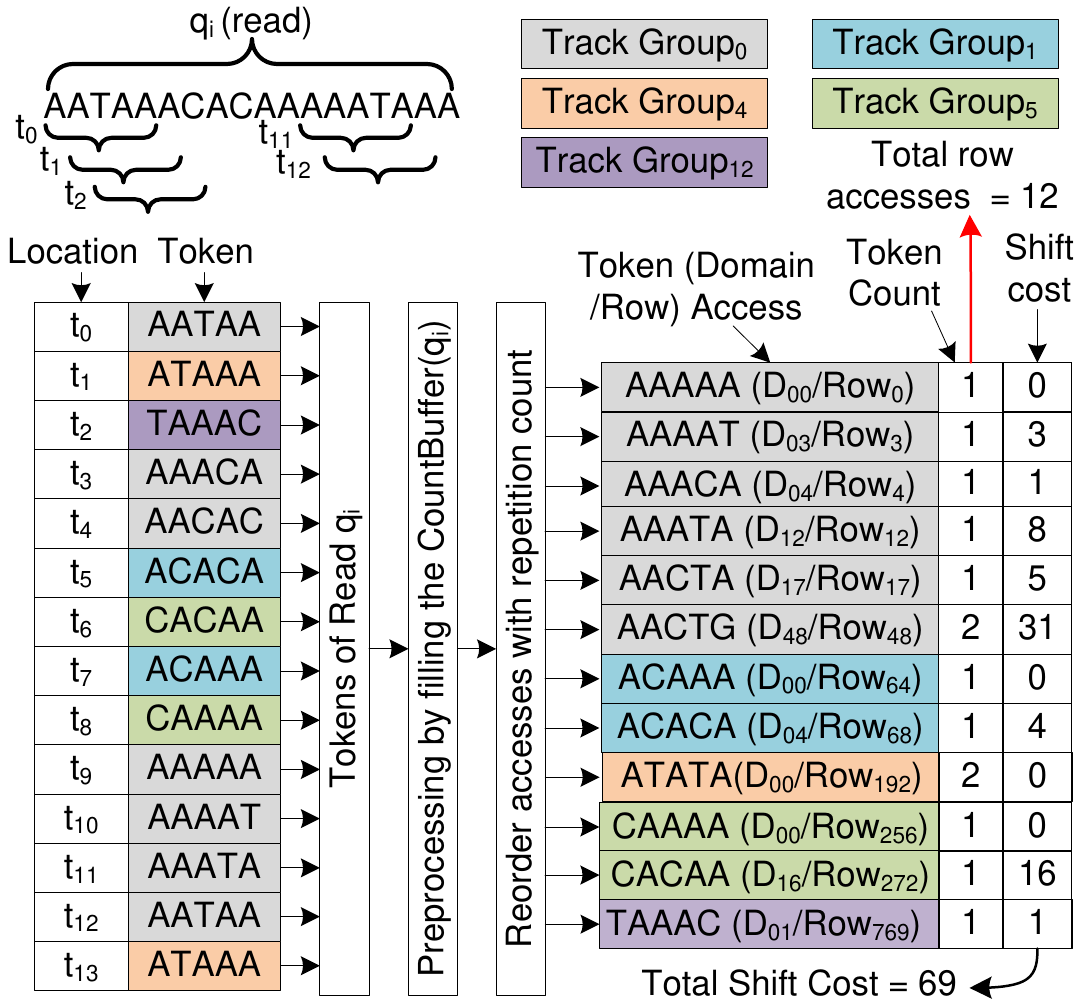}
\caption{ALPHA filter row access sequence for an example $q_i$ read assuming 64 domains per track.}
\vspace{-.1in}
\label{fig:ALPHA_Flow}
\end{figure}
\begin{figure}[bp]
\vspace{-.15in}
\centering
\includegraphics[width=.8\columnwidth]{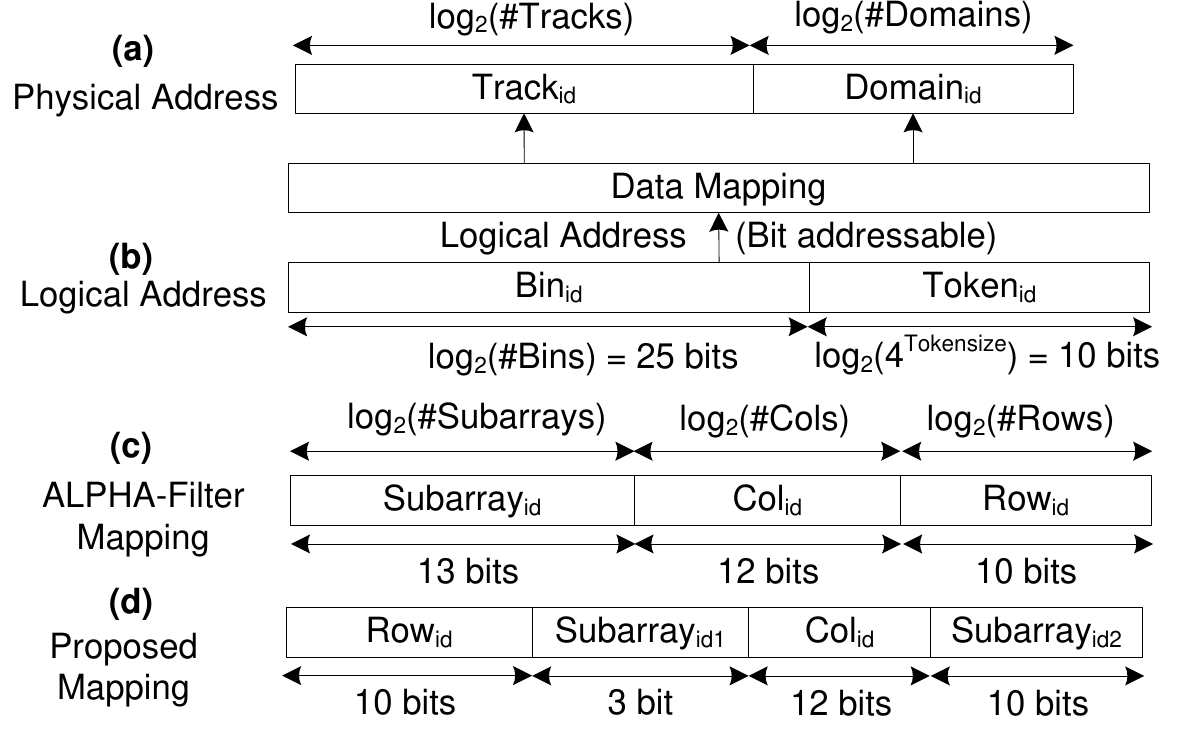}
\vspace{-.05in}
\caption{Logical to physical address mapping comparison for ALPHA and FIRM.}
\label{fig:PropMap}
\end{figure}

\vspace{.05in}

Assume the set of distinct tokens in a $q_i$ is represented by $\Theta$, \textit{i.e.,}  $\Theta(q_i) = \{\text{distinct tokens in }q_i\}$. ALPHA~\cite{ALPHA_2021} stores the count information for each $t \in \Theta$ in a small dedicated buffer called \textit{CountBuffer}. 
After populating the CountBuffer in the preprocessing step, $q_i$ is compared to the reference bins to compute the matching score. For a particular bin $r_k \in R$, the accumulated sum is computed as $\sum_{t \in \Theta(q_i)} \text{CountBuffer}(t)*\vec{r}_k[f(t)]$.  

Using the CountBuffer only rows with non-zero token counts 
are accessed as in Fig.~\ref{fig:ALPHA_Flow}. The first access is performed on Row$_{0}$ while the last access is performed on Row$_{769}$.  Row$_{48}$ and Row$_{192}$ are accessed only once, and $c_{k} +=2$  $\forall r_k \in R$ whose presence bits matching $AACTG$ and/or $ATATA$ are set.  The CountBuffer improvement can reduce RTM shifting to test $r_k$ in token index order rather than order of appearance in $q_i$.  However, the Countbuffer alone is insufficient to minimize RTM shifting, which we address in the next section.

\vspace{-.1in}
\vspace{-.05in}
\section{Shift and parallelism aware data mapping}
\label{sec:DataMap}
This section explains our proposed \emph{filtering in racetrack memory} methodology and architecture. FIRM implements DNA pre-alignment filtering~\cite{ALPHA_2021} (see Section~\ref{subsec:ALPHA}), using a novel data mapping customized to optimize data access patterns for RTM to maximize parallelism and minimize shifting overhead. 

\begin{figure*}[bth]
\centering
\begin{minipage}{.88\linewidth}
\includegraphics[width=\linewidth]{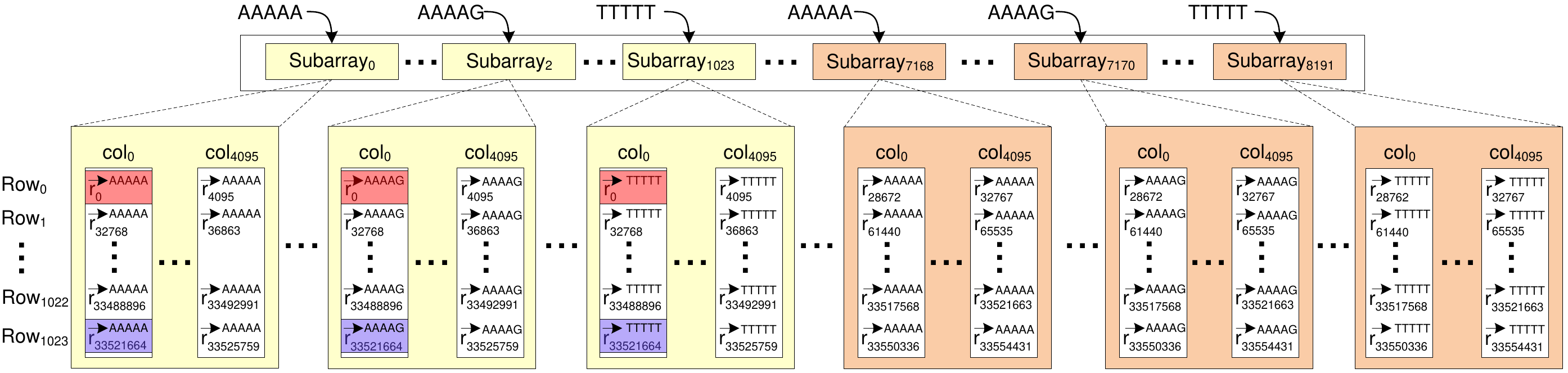}
\caption{Proposed data mapping from Fig.~\ref{fig:PropMap}d applied to the subarrays in Fig.~\ref{fig:GRIM3D}} 
\label{fig:PropMapExample}
\end{minipage}
\begin{minipage}{.115\linewidth}
\includegraphics[width=\linewidth]{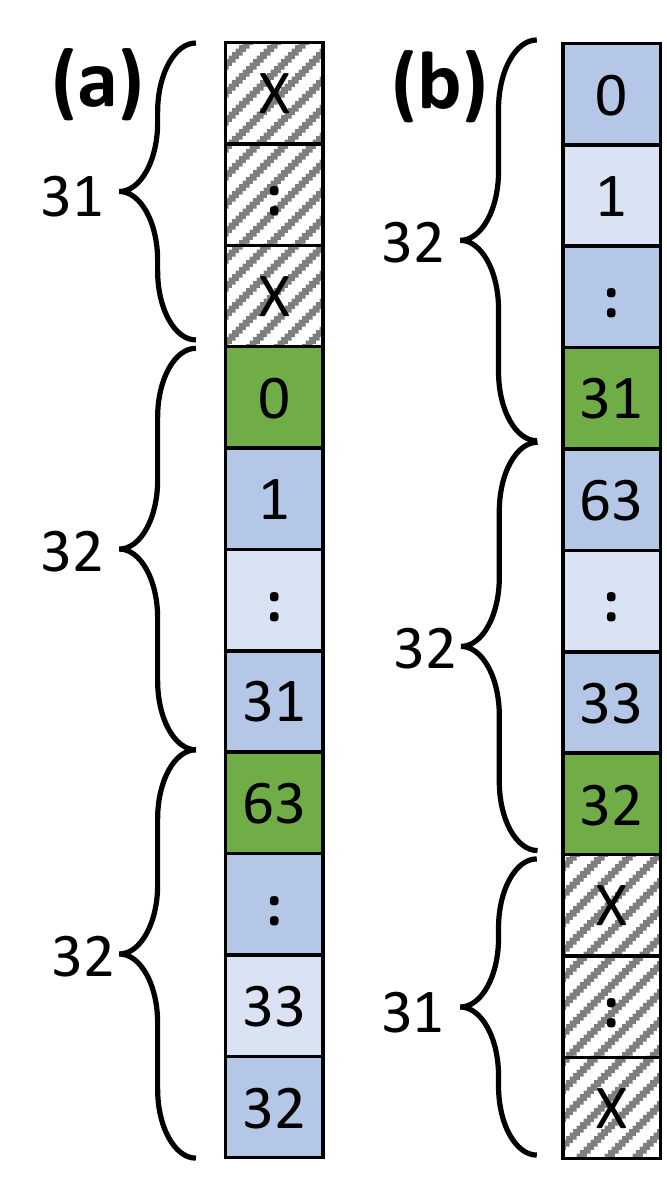}
\caption{US buf.}
\label{fig:US buf}
\end{minipage}
\vspace{-.1in}
\end{figure*}

FIRM defines which $\vec{r}_k[f(t)]$ is mapped to a particular subarray, track, and domain in RTM.  
For RTM, the logical structure of subarray, row, and column is extended into \textit{track groups} shown in Fig.~\ref{fig:GRIM3D}~\cite{pieee}. The track group is further partitioned by track and domain as a physical address as shown in Fig.~\ref{fig:PropMap}(a). 
Mapping 
the bins and tokens requires logically using the lower 10-bits for the token address and the upper bits for the bin address as in Fig.~\ref{fig:PropMap}(b).  In GRIM and ALPHA, the data mapping is done in a SUBARRAY-COL-ROW fashion for DRAM, as shown in Fig.~\ref{fig:PropMap}(c).  While this allows a column to represent a distinct bin, it requires subsequent tokens to be mapped into the same subarray which requires shifting and other memory activities such as activate and precharge to proceed sequentially.
Instead we remap the reference genome to RTM by interleaving the bit-vectors across subarrays, \textit{i.e.,} $\vec{r}_0[0]$ is mapped to Row$_0$ in Subarray$_0$, $\vec{r}_0[1]$ in Subarray$_1$ and $\vec{r}_0[1023]$ in Subarray$_{1023}$ 
so that the shifting overhead is minimized as in Fig.~\ref{fig:PropMap}(d).  This provides two important improvements in RTM.  First, memory accesses can be pipelined, mitigating the shift and other memory delays.  Additionally, as each Subarray stores the presence bit information of unique tokens, only Subarrays matching non-zero CountBuffer entries need to be accessed.

\begin{figure}[tbp]
\centering
\includegraphics[width=\linewidth]{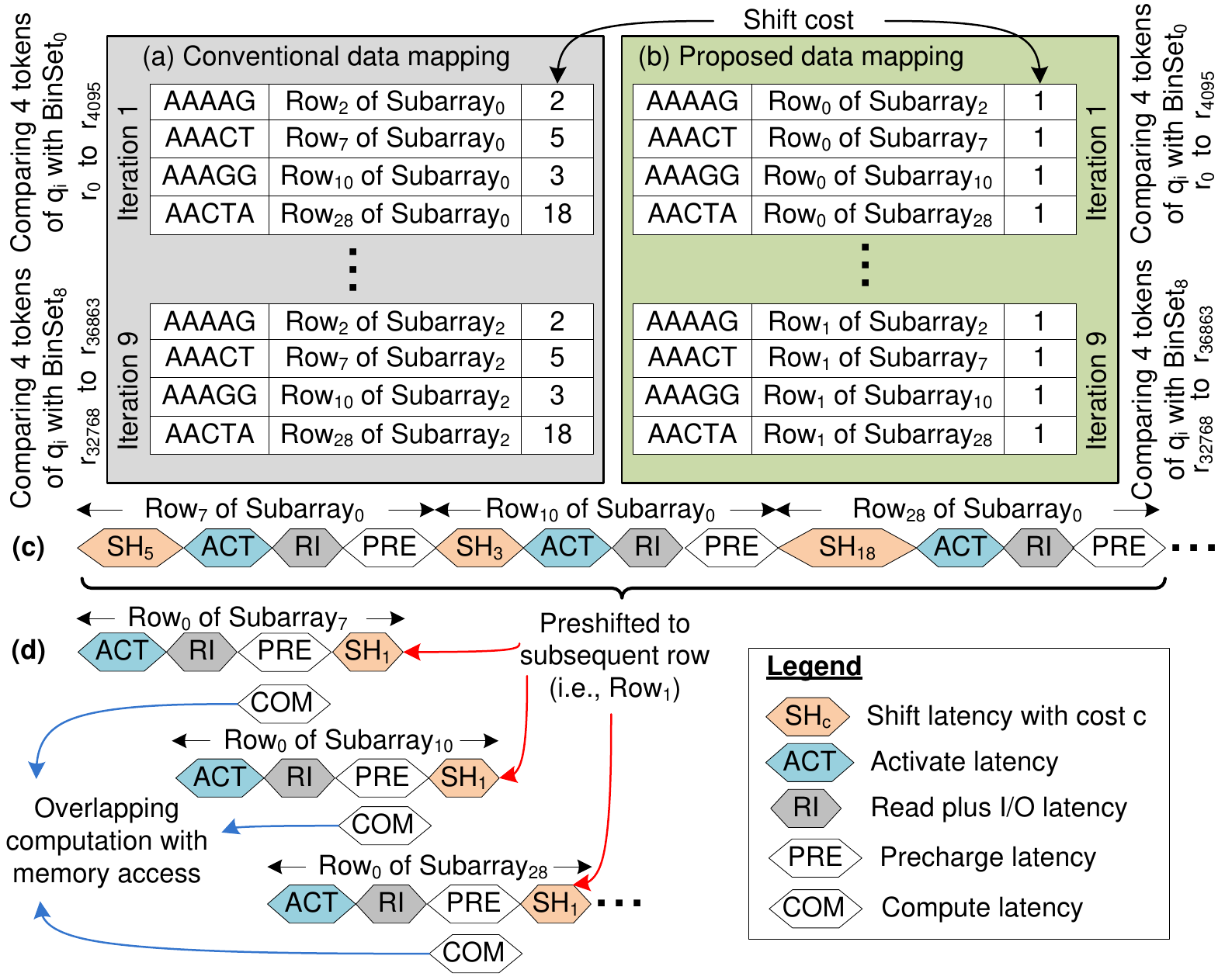}
\caption{Comparison of ALPHA and FIRM on shift count with details on memory pipelining potential for each approach.
} 
\label{fig:Comparison}
\end{figure}

Fig.~\ref{fig:PropMapExample} shows in detail how FIRM interleaves each token across different subarrays.  Consider a query read $q_i$ which contains $AAAAG$, $AAACT$, $AAAGG$, and $AACTA$ as its first four non-zero tokens. 
We presume that the number of bins that are compared in parallel with a particular read $q_i$ is referred to as a \emph{binset} which is equal to the row width. An \textit{iteration} is the processing of all bins within a binset with $q_i$.
Fig.~\ref{fig:Comparison}(a) describes the row accesses in a direct implementation of the ALPHA filter on RTM, which we note all map to the same Subarray (see Fig.\ref{fig:GRIM3D}).  We note that the shifting delays plus \texttt{ACT}, \texttt{RI}, and \texttt{PRE} times are sequential, shown in Fig.~\ref{fig:Comparison}(c). Thus, each iteration requires 28 shifts to process the first four tokens. In contrast, FIRM accesses Row$_0$ of the required Subarrays, which can pipeline accesses as shown in Fig.~\ref{fig:Comparison}(d) requiring only one shift per access as in Fig.~\ref{fig:Comparison}(b).  In fact, shifting is not required until the ninth iteration when the Subarrays, starting with Subarray$_2$, are revisited.
We use \textit{preshifting} to overlap the shift latency with the computation to reduce the impact of shift latency on the runtime of pre-alignment filtering. 
As soon as a row is accessed, the port positions in the active Subarray are shifted and aligned to the following row. 
Note, preshifting does not interfere with row accesses performed on independent subarrays and hides the shift latency because the row request is guaranteed to be serviced later.

Once $q_i$ is compared to the entire reference genome, the track groups in the shifted subarrays must to be reset to the first row to prepare for $q_{i+1}$. Note that, unlike the conventional mapping, in the proposed mapping, only those subarrays are accessed whose tokens have non-zero values in the CountBuffer (for a particular read). These subarrays are preshifted to the first row after accessing the last row in the track group to avoid any latency penalty. 

Although preshifting can effectively hide the track group reset delay, they still contribute to the RTM energy consumption. To overcome this, we can add a second access port to allow a circular, unlimited single shift buffer. As shown in Fig.~\ref{fig:US buf}, this allows each track group to avoid resetting shifts similar to prior work using RTM to store looped instructions~\cite{shrimp_TC_21}. In the \textit{reset} position (Fig.~\ref{fig:US buf}a) the track group starts with the data shifted down with local Row$_0$ aligned with access port as shown in green. The first half of the data is read by shifting up by one for each subsequent read until the track reaches the bottom position (Fig.~\ref{fig:US buf}b). The second half of the data is stored in reverse order so it can be accessed by the second access point shifting up between accesses. Once the last element is read, now the track group is back in the reset position.  
As a result, no additional reset shifts are required to prepare for $q_{i+1}$ cutting the required shifts in half. 
The additional port per track increases the energy consumption, but the energy savings due to shift reduction outweighs this overhead.

\label{sec:PropArch}

\vspace{-.15in}
\section{Evaluation}
\label{sec:results}
\vspace{-.025in}



\begin{table}[tb]
\vspace{0.1in}
\centering
\caption{The data set acquired from~\cite{igsr} with read size and token size of 100 and 5 base pairs respectively}
\vspace{-.05in}
\centering
\label{tab:applications}
\begin{threeparttable}
    \begin{tabular}{c|c|c}
    \toprule
    Benchmark & Read size & No. of reads \\
    \midrule
    ERR240726\_1 / ERR240726\_2\tnote{1} 	&  100 & 4031354 / 4389429 \\
    ERR240727\_1 / ERR240727\_2\tnote{1}	&  100 & 4082203 / 4013341 \\
    \bottomrule
    \end{tabular}
    \begin{tablenotes}
    \scriptsize
        \item[1] Sources: \url{http://ftp.1000genomes.ebi.ac.uk/vol1/ftp/phase3/data/}
    \end{tablenotes}
\end{threeparttable}
\vspace{-.25in}
\end{table}
%
%
\begin{table*}[tbp]
\caption{Energy and latency values of DRAM~\cite{DRAMSpec} RTM~\cite{destiny} and CMOS accelerator using 32 \SI{}{\nano\meter} technology and 1 GHz clock.}
\label{tab:EnergyLatValues}
\vspace{-.1in}
\begin{tabular}{l|c|l|c|l|c|c}
\hline\hline
\multicolumn{2}{c|}{\textbf{General Memory Parameters}} & \multicolumn{2}{c|}{\textbf{DRAM Parameters}}                          & \multicolumn{3}{c}{\textbf{RTM Parameters -- Track Length 64 domains}}                                                 \\\hline
Memory size         & \SI{4}{\giga\byte}                              & ACT and PRE [\SI{}{\pico\joule}]                & 1964                & Access Points per Track                               & 1                  & 2                  \\
Subarrays           & 8192                             & Access energy [\SI{}{\pico\joule}]/bit          & 1.25                & Background power [\SI{}{\milli\watt}]              & 193                 & 208                 \\
Rows per Subarray   & 1024                             & I/O energy [\SI{}{\pico\joule}]/bit             & 0.40                & Read energy [\SI{}{\pico\joule}]/bit            & 0.647               & 0.692               \\
Cols per Subarray   & 4096                             & Background power [\SI{}{\milli\watt}]           & 410                 & Shift energy [\SI{}{\pico\joule}]/bit           & 0.231               & 0.231               \\\hline
Tracks per DBC      & 512                              & $t_{RAS}$-$t_{RCD}$-$t_{RP}$-$t_{CAS}$-$t_{WR}$ [cycles] & 20-8-8-8-8 & $t_{RAS}$-$t_{RCD}$-$t_{RP}$-$t_{CAS}$-$t_{WR}$ [cycles] & \multicolumn{2}{c}{9-4-$2S$-4-4} \\\hline
\multicolumn{2}{c}{\textbf{Reference Genome}}          & \multicolumn{5}{c}{\textbf{CMOS Accelerator -- Accumulates 1 Row (4096 bins) per cycle}}                                           \\\hline
Bins in $R$         & 33,554,032   & Dynamic Energy [\SI{}{\pico\joule}]/bit         & 1785                & Leakage Power [\SI{}{\milli\watt}]              & \multicolumn{2}{c}{16.40}                \\\hline\hline
\end{tabular}\end{table*}

For evaluation, we use four pair-end short read genomes from the 1000 Genome Project~\cite{igsr}.
We assume a $100$-nucleotid query and reference bins, with a token size of 5. 
Memory (DRAM and RTM) and accelerator architecture parameters including latency, and energy/power parameters are listed in Table~\ref{tab:EnergyLatValues} using DRAMSpec~\cite{DRAMSpec} for DRAM, DESTINY~\cite{destiny} for RTM, and Cadence RTL-Compiler Synthesis.  
RTM is evaluated for 64 domains per track
For system-level shifts, latency and energy estimation we use RTSim~\cite{rtsim}, an RTM simulator extended from NVMain~\cite{nvmain2.0}. 
We compare the following systems in our evaluation: 

\begin{itemize}
\vspace{-.06in}
\item \emph{GRIM:} The DRAM based GRIM filter~\cite{GRIM_2018}. 
shown in Fig.~\ref{fig:GRIM3D}. 
\item \emph{ALPHA:} The recently proposed DRAM based ALPHA filter design with preprocessing~\cite{ALPHA_2021}, cf. Section~\ref{subsec:ALPHA}.
\item \emph{ALPHA-RTM:} The ALPHA filter design applied to RTM. 
\item \emph{FIRM:} The FIRM filter, cf. Section~\ref{sec:PropArch}.
\item \emph{FIRMPR:} FIRM extended with the preshifting. 
\item \emph{FIRMUS:} FIRMPR with circular unlimited single-shifts.
\vspace{-.05in}
\end{itemize}

\begin{figure}[tb]
\vspace{-.1in}
\centering
\vspace{-.1in}
\includegraphics[width=\columnwidth]{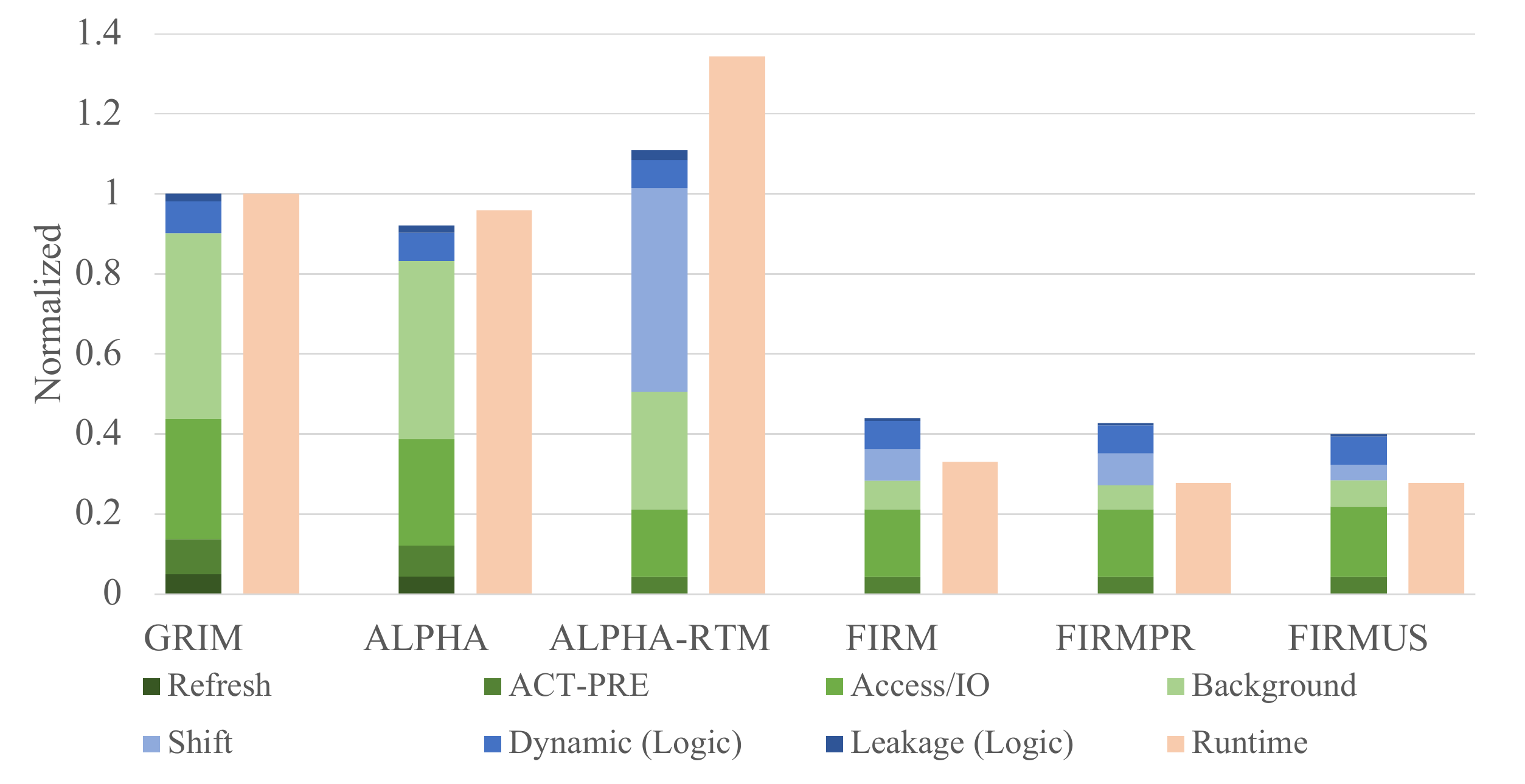}
\vspace{-.2in}
\caption{Normalized runtime and energy breakdown.} 
\label{fig:ShiftsRuntime}
\vspace{-.1in}
\end{figure}

Fig.~\ref{fig:ShiftsRuntime} compares the energy and runtime of different prefiltering solutions normalized to GRIM. 
Just replacing ALPHA's DRAM with RTM increases runtime and energy dominated by shift cost.  FIRM's 
exploiting subarray parallelism~\cite{SALP} replaces a single shift for frequent long shifts when successive RTM rows are accessed, reducing shifts by 84\% over ALPHA-RTM, reducing average runtime over ALPHA-RTM by 75\% and over ALPHA by 63\%.
This helps FIRM 
provide an energy savings of 43\% over ALPHA when combined with intrinsically reduced RTM background power.

Applying preshifting (FIRMPR) further improves the average runtime by 5.3\% compared to FIRM.
Applying the unlimited single-shift technique (FIRMUS) achieves the same runtime while reducing overall energy by 2.8\% over FIRMPR.  FIRMUS energy savings are attenuated by adding a second access port.

\vspace{-.18in}
\section{Conclusions}
\label{sec:conclusion}
\vspace{-.025in}

In this paper, we explore a die-stacked RTM design for implementing a pre-alignment seed location filtering algorithm. We motivate our design through an example that shows why existing optimizations for DRAM-based systems are not directly applicable to RTM-based designs.
We propose a novel filtering in Racetrack memory scheme that provides novel data layout, preshifting, and circular buffers to significantly reduce shift operations in RTM. Our experimental evaluations show that our proposal improves performance by more than $3\times$ while reducing energy to less than half of the state of the art appraoch.

\vspace{-.18in}
\section*{Acknowledgments}
This work was partially funded by the German Research Council DFG (projects 437232907, 366764507, 450944241) and US NSF (projects 1822085, 2133267) LPS, and NSA.  
\ifCLASSOPTIONcaptionsoff
  \newpage
\fi
\vspace{-.15in}
\bibliographystyle{IEEEtran}
\vspace{-.025in}
\bibliography{gogal}

\end{document}